\documentclass[a4paper]{jpconf}

\newcommand{\nin}{\noindent}
\newcommand{\be}{\begin{equation}}
\newcommand{\ee}{\end{equation}}
\newcommand{\bea}{\begin{eqnarray}}
\newcommand{\eea}{\end{eqnarray}}
\newcommand{\nn}{\nonumber\\}

\newcommand{\ap}{\alpha'}

\usepackage{graphicx}

\begin{document}
\title{Tachyon - Dilaton driven Inflation as an $\ap$ - non-perturbative solution in First Quantized String Cosmology}

\author{Anna Kostouki}

\address{King's College London, Dept. of Physics, Strand, London, WC2R 2LS, UK}

\ead{anna.kostouki@kcl.ac.uk}

\begin{abstract}
Applying a novel non-perturbative functional method framework to a two-dimensional bosonic sigma model with tachyon, dilaton and graviton backgrounds we construct exact (non perturbative in the Regge slope $\ap$) inflationary solutions, consistent with world-sheet Weyl Invariance. The mechanism for inflation entails a (partial) ``alignment'' between tachyon and dilaton backgrounds in the solution space. Some cosmological solutions which contain inflationary eras for a short period and interpolate between flat universes in the far past and far future are also discussed. These solutions are characterized by the absence of cosmological horizons, and therefore have well-defined scattering amplitudes. This makes them compatible with a perturbative string framework, and therefore it is these solutions that we consider as self-consistent in our approach. Within the context of the interpolating solutions, string production at the end of inflation (preheating) may also be studied. The advantage of our method is that the solutions are valid directly in four target-space-time dimensions, as a result of the non trivial dilaton configurations. Whether the model is phenomenologically realistic, with respect to its particle physics aspects, remains an open issue. This talk was based on work recently done with J. Alexandre and N. E. Mavromatos \cite{akm}.
\end{abstract}

\section{Motivation: String Inflation in $4$ Dimensions}

An inflationary era (i.e. an era of accelerating expansion) in the early universe can explain many cosmological observations, such as the ``horizon problem'' or the large-scale structure that is observed in the sky today. The initial single-field inflation \cite{inflation} is a particularly elegant and simple model. Inflation can be viewed either as a fundamental theory or as an effective theory of some underlying theory of quantum gravity. The latter is the case in String Theory. A lot of models for inflation exist in the context of String theory, but problems such as the need of compactification of the extra dimensions that String Theory introduces arise. In general inflationary models coming from String Theory, fail to be simple and elegant, like the original inflationary model.

In non-critical String Theory there is no need for extra dimensions. The introduction, for example, of a dilaton background field changes the conformal properties of Bosonic String Theory in such a way that a model can be viable in $4$ dimensions. In this work we study a model for inflation coming from a bosonic closed string with a graviton, dilaton and tachyon background, that can be realized in $4$ dimensions, and is consistent with conformal invariance conditions, as a resummation to all orders in $\ap$. The latter is important for early universe String Cosmology, where higher order curvature terms play a significant role. These terms may appear as corrections to the Einstein action (which in String Theory is viewed as an effective action for the background fields in the target space-time), to second or higher order in $\ap$. Therefore, a perturbative in $\ap$ treatment is probably not sufficient for early universe String Cosmology.

Closed string tachyon condensation is a subject which has been studied widely in the literature \cite{Hellerman,swanson,Silverstein,Horava,Frey} (mainly in the context of Superstring or Heterotic string theory), as it seems to be closely related to the study of cosmological singularities. However, these studies don't go further than a perturbative in $\ap$ treatment of the problem, and therefore find solutions connected to the perturbative vacuum, always assuming for example a linear solution for the dilaton field. It is also interesting to note that some exact solutions have been found \cite{swanson,Frey}. The difference with our work is that these solutions are also space-dependent, whereas we are assuming only time-dependence of the backgrounds, and that in these works a weak tachyon field is assumed, whereas in our solutions the tachyon field may take values much greater than $T=0$.

\section{CFT on the world-sheet}

We consider a closed bosonic string, living in a $D$-dimensional target space-time, with a graviton, $g_{\mu\nu}$, dilaton, $\phi$, and tachyon, $T$, background. Because we are interested in the cosmological properties of this model, we assume these background fields to be dependent only on the time coordinate of the string, $X^0$. The two-dimensional quantum theory on a world-sheet with metric $\gamma_{ab}$ and curvature $R^{(2)}$ is described by the action:
\be
S=\frac{1}{4\pi\ap} \int d^2\sigma \sqrt{\gamma} \left[g_{\mu\nu}(X^0)\partial_a X^\mu \partial_b X^\nu + \ap R^{(2)} \phi(X^0) + \ap T(X^0) \right]
\ee

In a way that lies beyond the limits of this talk, following a non-perturbative (in $\ap$) field theoretical method and previous work \cite{AEM1}, we chose the following time-dependent configuration for the three background fields:
\bea\label{config}
g_{\mu\nu}&=&\frac{A}{(X^0)^2}\eta_{\mu\nu}\nn
\phi&=&\phi_0\ln\left(\frac{X^0}{\sqrt{\alpha'}}\right)\nn
T&=&\tau_0\ln\left(\frac{X^0}{\sqrt{\alpha'}}\right).
\eea
\nin where $\phi_0$ and $\tau_0$ are dimensionless constants, and $A$ is a constant with dimensions $[\mbox{mass}]^{-2}$. We then proceed to study the conformal properties of the model. Conformal invariance is a very important property of the theory that has to be maintained when quantizing it. This is ensured as long as the Weyl anomaly coefficients of the model, $\beta^{g}_{\mu\nu}$, $\beta^\phi$ and $\beta^T$ are equal to $0$. The Weyl anomaly coefficients have the following form, to lowest order in $\ap$:
\bea\label{1loop}
\beta^g_{\mu\nu}&=&\alpha'R_{\mu\nu}+2\alpha'\nabla_\mu\nabla_\nu\phi
-\alpha^{'}\partial_\mu T\partial_\nu T+{\cal O}(\alpha')^2\nn
\beta^\phi&=&\frac{D-26}{6}-\frac{\alpha'}{2}\nabla^2\phi+\alpha'\partial^\mu\phi\partial_\mu\phi+{\cal O}(\alpha')^2\nn
\beta^T&=&-2T-\frac{\alpha'}{2}\nabla^2 T+\alpha'\partial^\mu\phi\partial_\mu T+{\cal O}(\alpha')^2,
\eea
We argue now that one can always find a renormalization group scheme in which $\beta^{g}_{\mu\nu}$, $\beta^\phi$ and $\beta^T$ vanish at arbitrary order in the $\ap$-expansion. This will be based on the freedom to redefine the background fields without affecting the scattering amplitudes of the theory.
When plugging into the above expressions the configuration (\ref{config}), we see that the $1$-loop Weyl anomaly coefficients do not vanish. However, we argue that the resummed to all orders in $\ap$ Weyl anomaly coefficients do so, due to the following observation: This configuration leads to $\beta^{g}_{\mu\nu}$, $\beta^\phi$ and $\beta^T$, to first order in $\ap$, that have a homogeneous dependence on $X^0$ (e.g. in $\beta^{g}_{\mu\nu}$ all terms are proportional to $(X^0)^{-2}$), besides one term in $\beta^T$, which is linear in $T\propto \ln X^0$:
\bea\label{betas}
\beta^g_{00}&=&-\frac{\alpha^{'}}{(X^0)^2}(D-1+\tau_0^2)+{\cal O}(\ap^2)\nn
\beta^g_{ij}&=&\frac{\alpha^{'}\delta_{ij}}{(X^0)^2}(D-1+2\phi_0)+{\cal O}(\ap^2)\nn
\beta^\phi&=&\frac{D-26}{6}+\frac{\alpha^{'}}{2}\left(D-1+2\phi_0\right)\frac{\phi_0}{A}+{\cal O}(\ap^2)\nn
\beta^T&=&-2T+\frac{\alpha^{'}}{2}\left(D-1+2\phi_0\right)\frac{\tau_0}{A}+{\cal O}(\ap^2)
\eea
Based on power counting, we also know that any terms that may appear in the Weyl anomaly coefficients at higher orders in $\ap$, will be homogeneous to the above ones. We now use the freedom to make general field redefinitions, $g^i \rightarrow \tilde g^i$  ($g^i=g_{\mu\nu},\phi,T$), that leave the theory invariant, but under which the Weyl anomaly coefficients transform in the following way \cite{metsaev}:
\be
\beta^i \rightarrow \tilde \beta^i = \beta^i + \delta g^j \frac{\delta \beta^i}{\delta g^j}-\beta^j \frac{\delta(\tilde g^i - g^i)}{\delta g^j}
\ee
Using this freedom we can make the Weyl anomaly coefficients vanish order to order, in all orders, starting from their value in second order in $\ap$. We then employ a specific field redefinition that contains $19$ free parameters and doesn't change to dependence of the fields on $X^0$. It is beyond the scope of this talk to give many details on the form and on the effect of this redefinition. There are two important facts about it: The first one is that there is one ``special'' terms in the redefinition of the tachyon field (namely one that is proportional to $\ap R T$, where $R$ is the target space-time curvature), in the sense that it causes the appearance of a new linear term in $\beta^T$. Fixing the value of one of the free parameters, we can make this term cancel the original $-2T$ and thus be left with only homogeneous terms in all the Weyl anomaly coefficients. The second important fact is that every other term in the field redefinitions causes the appearance of new terms, that are of second order in $\ap$, and are homogeneous to the old ones. Thus, we end up with something like:
\bea
\beta^{g}_{\mu\nu}&=&\frac{E_{\mu\nu}}{(X^0)^2}+\cal O \left(\ap^2 \right) \nn
\beta^\phi &=& E_1 +\cal O \left(\ap^2 \right) \nn
\beta^T &=& E_2 +\cal O \left(\ap^2 \right)
\eea
\nin where $E_{\mu\nu}$, $E_1$ and $E_2$ are constants that are linear combinations of the parameters that we introduced in the field redefinition. Fixing the value of (some of) the $18$ free parameters left, we have enough freedom to impose $E_{\mu\nu}=E_1=E_2=0$ and thus make the Weyl anomaly coefficients vanish to second order in $\ap$. This procedure can be repeated order by order, to all orders in $\ap$. Thus, we claim that our configuration (\ref{config}) satisfies conformal invariance conditions to all orders in $\ap$.

\section{Cosmology of the model}

After checking the conformal properties of our configuration (\ref{config}), we now proceed to study the resulting cosmology. In order to do this, we need to study the effective theory in the target space-time of our world-sheet theory. It is again the conformal invariance conditions of the theory on the worldsheet, $\beta^{g}_{\mu\nu}=\beta^\phi=\beta^T=0$, that define the effective target-space theory (notice that these conditions, e.g. equations (\ref{1loop}),when written out explicitly, look like equations of motion for the background fields). However, the exact form of an effective action to describe tachyon backgrounds of closed strings is not known. We therefore choose to work with the most general two-derivative action for a closed string with a graviton, dilaton and tachyon background \cite{swanson}:
\bea \label{action}
S^{(D)}&=& \int d^D x \sqrt{-g} e^{-2\phi} \Biggl\{ \frac{D-26}{6\ap} + f_0(T) +f_1(T) R + 4 f_2(T) \partial_\mu\phi \partial^\mu \phi \nn
&&~~~~~~~~~~~~~~~~~~~~~~ - f_3(T) \partial_\mu T \partial^\mu T - f_4(T) \partial_\mu T \partial^\mu \phi \Biggr\}
\eea
\nin where $d^Dx=dx^0 d{\bf x}$ and $x^\mu$ denotes the zero mode of $X^\mu$. Note that the choice of the functions $f_i(T)$ is not unique. Field redefinitions, that as we mentioned above leave the scattering matrix of the theory invariant, will lead to different forms of these functions. In the following, we will try to make some assumptions about the form of these functions and put some constraints on them.
The effective action (\ref{action}) is written in the ``sigma-model frame", where the Einstein-Hilbert term in the action doesn't take its canonical form, $\int d^Dx \sqrt{-g} R$, but instead is $\int d^Dx \sqrt{-g} e^{-2\phi}f_1(T)R$. By doing the following metric redefinition,
\bea\label{pass}
&&g_{\mu\nu}\rightarrow g^{E}_{\mu\nu}=e^{\omega(\phi,T)} g_{\mu\nu} \nn
&&~~~~\omega(\phi,T)=\frac{-4\phi+2\ln f_1(T)}{D-2}~,
\eea
we pass to the ``Einstein frame", where the Einstein-Hilbert term takes its canonical form. Our metric configuration (in the sigma-model frame) is $g_{\mu\nu}=\frac{A}{(x^0)^2}\eta_{\mu\nu}$. This is an inflationary metric in the sigma-model frame:
\be
ds^2=\frac{A}{(x^0)^2} \left[(dx^0)^2-(d{\bf x}))^2  \right]
\ee
After making the coordinate redefinition $x^0\rightarrow y^0=-\sqrt{A}\ln \frac{x^0}{\ap}$, $x^i \rightarrow y^i = \sqrt\frac{A}{\ap}x^i$,
\be
ds^2=(dy^0)^2-e^{2y^0/\sqrt{A}}(d{\bf y})^2 ~,
\ee
\nin we clearly see that the scale factor grows exponentially with time, $a(y^0)=e^{y^0/\sqrt{A}}$, and the Hubble rate is equal to $H=\frac{1}{\sqrt{A}}$.
When we pass to the Einstein frame, the metric becomes
\be
ds^2\equiv dt^2-a^2(t) d{\bf r}^2 = \frac{A e^\omega}{(x^0)^2} \left[(dx^0)^2-(d{\bf x}))^2   \right]
\ee
From the above expression, one can in principle find the relation between the cosmic time, $t$, and the sigma-model frame time, $x^0$, and a precise expression for the scale factor in the Einstein frame, $a(t)$, as long as one knows the form of the function $f_1(T)$. The form of the configuration for the dilaton and the tachyon fields is also important in the calculation of $t(x^0)$ and $a(t)$. In our case, one should always keep in mind that we have $\phi=\phi_0\ln \frac{x^0}{\sqrt{\ap}}$ and $T=\tau_0\ln \frac{x^0}{\sqrt{\ap}}$.

\subsection{Eternal inflation}
In ref. \cite{tseytlin} the simple choice $f_1(T)=e^{-T}$ is made, based on a definition of the target-space effective action in the sigma-model frame that relates it to the world-sheet partition function, $Z$:
\be
S^{(D)}=\beta^i\cdot \frac{\delta}{\delta g^i}Z
\ee
\nin However, this choice is dismissed further on in the same reference, as it leads to an action which doesn't contain the standard perturbative closed bosonic string vacuum ($D=26$, $\phi=T=0$ and $g_{\mu\nu}=\eta_{\mu\nu}$). This is a well-known problem associated with the non-trivial tachyon tadpoles. Our approach is a bit different: the tachyon-tadpole created the unwanted inhomogeneity in $\beta^T$ that we discussed in the previous section. However, we have removed this by making a specific choice for a field redefinition. Furthermore, the perturbative vacuum doesn't need to be connected to our configuration, which is a solution to the conformal invariance conditions of the world-sheet theory that holds to all orders in $\ap$. Therefore, we assume that the choice $f_1(T)=e^{-T}$ is valid within our approach and we proceed to study the effective space-time theory with this choice.
The function $\omega(\phi,T)$ that appears in the metric redefinition (\ref{pass}) takes the following form for the choice $f_1(T)=e^{-T}$:
\be
\omega(\phi,T)=-\frac{4\phi+2T}{D-2}=-\frac{4\phi_0+2\tau_0}{D-2} \ln x^0
\ee
\nin We note immediately that if the amplitudes of the two fields satisfy an ``anti-alignment" condition,
\be \label{antial}
2\phi_0+\tau_0=0~,
\ee
\nin the Einstein frame coincides with the sigma-model frame. Thus, we have inflation in both frames, with the Hubble rate in the Einstein frame equal to $\frac{1}{\sqrt{A}}$ as well. The cosmic time, $t$, in this case is related to $x^0$ by:
\be\label{time1}
t\propto-\sqrt{A}\ln\frac{x^0}{\sqrt{\ap}}
\ee
\nin and the scale factor, $a(t)$, grows exponentially with time:
\be
a(t)=a_0 \exp \left(\frac{t}{\sqrt{A}} \right)
\ee
If the anti-alignment condition (\ref{antial}) is not satisfied, in the Einstein frame we have a power-law expanding universe. The cosmic time is given by
\be\label{time2}
t\propto \left(\frac{x^0}{\sqrt{\ap}} \right)^{-\frac{2\phi_0+\tau_0}{D-2}}~,
\ee
\nin and the scale factor grows as a power of $t$:
\be
a(t) \propto t^{1+\frac{D-2}{2}\left(\phi_0+\frac{\tau_0}{2}\right)^{-1}}~.
\ee
In this case, the condition
\be\label{cond}
2\phi_0+\tau_0 >0
\ee
\nin is sufficient to guarantee that the expanding universe is not characterized by cosmic horizons, and thus the string scattering matrices are well defined.

\subsection{Exit from inflation}\label{exit}
So far we saw that an eternally inflating universe is possible within our configuration. However, this inflationary era is only possible when the anti-alignment condition (\ref{antial}) holds. Is there a way to exit from this eternal inflation era, and enter a power-law expansion (or even Minkowski era)? One way would be to disturb the anti-alignment condition. For example, this could be done through a (yet unknown) mechanism that would make the tachyon field decay to zero. Another option is to make a different choice for the function $f_1(T)$. Different choices of $f_1(T)$ lead to different solutions of the Weyl invariance conditions, that should be related to each other through local field redefinitions. Field redefinitions leave the scattering amplitudes invariant, so the latter statement is only valid in cases where perturbative String Theory is well-defined. This is not the case for a de Sitter universe, where horizons are unavoidable. In the following, we exploit solutions that correspond to choices for $f_1(T)$ (that are resummations of exponentials of $T$), which cannot emerge from a pure de Sitter through a field redefinition, but contain inflationary eras for certain periods and interpolate between flat universes. Horizons are not an issue in such a case, and perturbative String Theory is well-defined. Since our approach is based on scattering amplitudes (these should in principle determine the functions $f_i(T)$), it is these solutions that we consider to be self-consistent.

\begin{figure}[th]
\centering
\includegraphics[width=12.6cm, height=8.4cm]{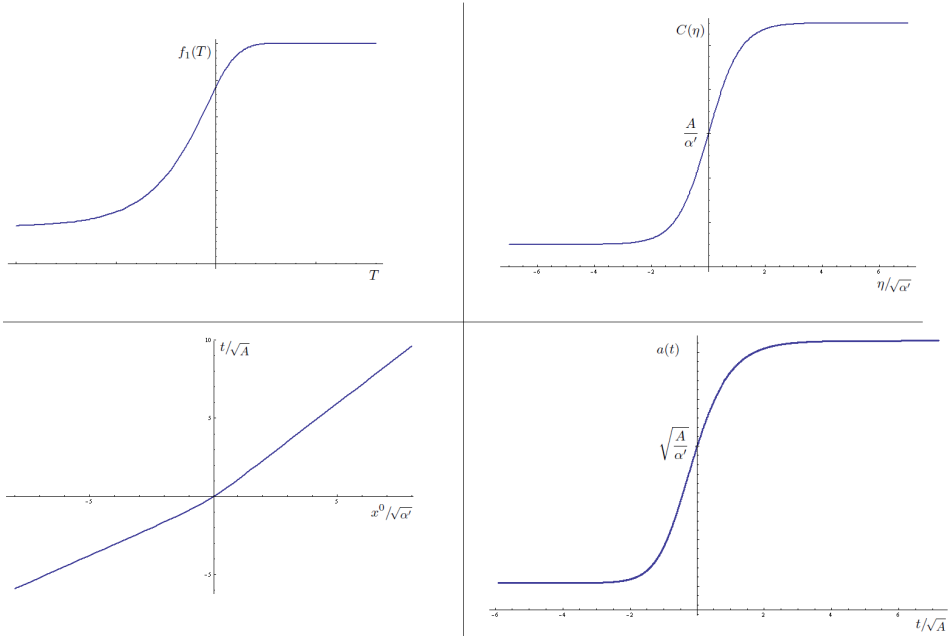}
\caption{{\footnotesize {\it {Upper left}}: Plot of the function $f_1(T)=1+0.9\tanh(e^{T})$. For dilaton and tachyon amplitudes satisfying $\phi_0=-1$ and $\tau_0=1$, this leads to the conformal scale factor plotted in the {\it{upper right}} part of this figure. {\it Lower left}: Plot of the flow of the cosmic time $t$ w.r.t. the sigma-model frame time, $x^0$. {\it Lower right}: Plot of the scale factor $a(t)$ in this case.}}
\label{fig1}
\end{figure}

If we go back to the case of a general function $f_1(T)$, we can rewrite the metric in the Einstein frame, taking into account the specific configurations for the dilaton and the tachyon fields, as:
\bea
ds^2&=&dt^2-a^2(t)d{\bf r}^2 \equiv C(\eta) \left(d\eta^2-d{\bf r}^2\right) \nn
&=& \frac{A}{\ap} \left(\frac{x^0}{\sqrt{\ap}}\right)^{-2(1+\phi_0)}f_1\left( \tau_0 \ln \frac{x^0}{\sqrt{\ap}}\right) \Bigl[(dx^0)^2-d{\bf x}^2 \Bigr]~.
\eea
\nin It is clear from the above expression that $x^0$ plays the role of conformal time, $\eta$, (in both frames) independently of the form of $f_1$. The form of the conformal scale factor, $C(\eta)$, arises straightforwardly from $f_1(T)$:
\be
C(\eta)=\frac{A}{\ap} \left(\frac{\eta}{\sqrt\ap}\right)^{-2(1+\phi_0)} f_1\left( \tau_0 \ln \frac{\eta}{\sqrt{\ap}}\right)
\ee
Using this result, we will study two classes of functions $f_1(T)$ that lead to interesting cosmologies, as they interpolate between Minkowski universes, but contain inflationary eras. Particle (or string) production at the end of the inflationary era can also be discussed \cite{gubser} in these cases. The first of them is functions of the form
\be\label{f1}
f_1(T)=1+B\tanh \left(\sqrt \ap \rho e^{T/\tau} \right)~,
\ee
\nin where $B$, $\tau$ and $\rho$ are constants. A function of this type is plotted in figure \ref{fig1}. If the dilaton and tachyon amplitudes satisfy the conditions $\phi_0=-1$ and $\tau_0=\tau$, then this leads to a conformal scale factor of the form
\be\label{c1}
C(\eta)=\frac{A}{\ap} \left[1+B\tanh(\rho\eta) \right]~,
\ee
\nin that is plotted in figure \ref{fig1} as well. The corresponding scale factor $a(t)$ is also plotted there. This is a cosmological model which interpolates between two flat epochs and contains an inflationary era for a certain time period. We get similar results for the class of functions $f_1(T)$:
\be\label{f2}
f_1(T)=1+B\frac{e^{T/\tau}}{\sqrt{e^{2T/\tau}+\rho^2/\ap}}~.
\ee
If $\phi_0=-1$ and $\tau_0=\tau$, the conformal scale factor is given by:
\be\label{c2}
C(\eta)=\frac{A}{\ap} \left(1+B\frac{\eta}{\eta^2+\rho^2} \right)~.
\ee
Particle production at the end of inflation has been studied in \cite{gubser} for the above two scale factors, (\ref{c1}) and (\ref{c2}). As already argued above, since these solutions interpolate between flat universes, with well-defined S-matrix elements, these will be our choice as self-consistent configurations.

\begin{figure}[th]
\centering
\includegraphics[width=12.6cm, height=8.4cm]{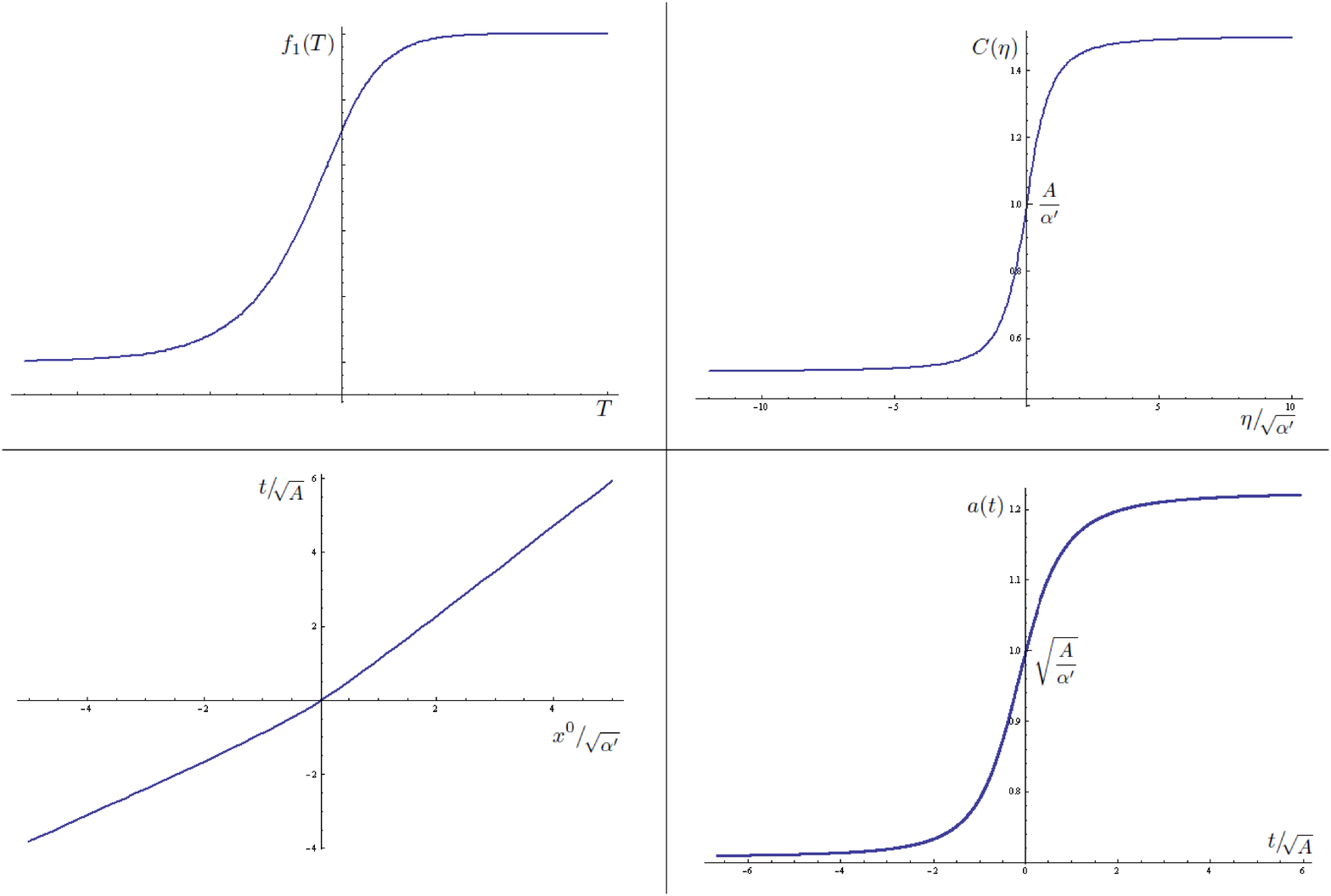}
\caption{{\footnotesize {\it {Upper left}}: Plot of the function $f_1(T)=1+0.5\frac{e^T}{\sqrt{e^{2T+\frac{1}{\ap}}}}$. For dilaton and tachyon amplitudes satisfying $\phi_0=-1$ and $\tau_0=1$, this leads to the conformal scale factor plotted in the {\it{upper right}} part of this figure. {\it Lower left}: Plot of the flow of the cosmic time $t$ w.r.t. the sigma-model frame time, $x^0$. {\it Lower right}: Plot of the scale factor $a(t)$ in this case.}}
\label{fig2}
\end{figure}

\subsection{Consistency and stability checks of our solutions}
In a universe that contains tachyonic fields, some questions about stability may arise. However, we start by checking something that one may question because of the presence of a dilaton field. The string coupling needs to be much less than one asymptotically, for $t\rightarrow \infty$, in order to ensure perturbative validity of our string tree-level considerations. The string coupling is related to the value of the dilaton field, as $g_s\propto \exp(\phi)$. We will check that this falls asymptotically to zero for all of the above studied solutions. For the solutions for our first choice of $f_1$, $f_1(T)=e^{-T}$, from the relations between the cosmic time $t$ and $x^0$, (\ref{time1}) and (\ref{time2}), we see that
\bea
g_s & \propto &-\phi_0 \frac{t}{\sqrt{A}}~~,~~~~~~~~~~~\mbox{if}~~2\phi_0+\tau_0=0 \nn
& \propto & \left(\frac{t}{\sqrt{A}}\right)^{-\frac{D-2}{2\phi_0+\tau_0}\phi_0}~,~~\mbox{if}~~2\phi_0+\tau_0\neq0
\eea
From the above expressions, we deduce that we have to place the following constraint on the dilaton tachyon amplitude (taking also into account the constraint (\ref{cond}) which prevents the existence of cosmic horizons), in order to ensure that the string coupling falls to zero as $t\rightarrow \infty$:
\be
\phi_0>0
\ee
For the second and third choices of $f_1$, (\ref{f1}) and (\ref{f2}), we note from figures \ref{fig1} and \ref{fig2} that in both cases, for $x^0\rightarrow\infty$, $t$ behaves like $t\propto x^0$. This means that $\phi\propto\phi_0\ln\frac{t}{\sqrt{\ap}}=-\ln\frac{t}{\sqrt\ap}$, which is enough to ensure that
\be
g_s\propto t^{-1}~~\Rightarrow~~g_s\rightarrow 0 ~~\mbox{when}~t\rightarrow \infty~.
\ee

Another question that usually arises in string models with dilaton and tachyon backgrounds is whether there are any "ghost" fields, which means fields whose kinetic terms appear with the wrong sign in the (effective) target-space action. In our case, in the action (\ref{action}), the dilaton field appears initially with the wrong sign (we assume for all the functions $f_i(T)$ that $f_i(T)=1+{\cal O} (T)$). To check, however, if there are really any ghost fields, we need to use the complete Einstein-frame effective action and diagonalize it, by redefining $\left(\phi,T\right)\rightarrow\left(\tilde\phi,\tilde T\right)$, in order to be left only with $\partial^\mu\tilde\phi\partial_\mu\tilde\phi$ and $\partial^\mu \tilde T\partial_\mu \tilde T$ terms. After diagonalizing the action, we can place some - mild - constraints on the functions $f_i(T)$ in order to ensure absence of ghost fields. In the most general case, these are:
\bea \label{restrictions}
&& f_2+\frac{\left[4(D-1)f_1'-(D-2)f_4 \right]^2}{8(D-2)\left[(D-1)\frac{(f_1')^2}{f_1}+(D-2)f_3\right]} ~<~ \frac{D-1}{D-2} f_1 \nn
&& f_3+\frac{D-1}{D-2}\frac{(f_1')^2}{f_1} ~>~0
\eea
\nin For the last two choices that we made for $f_1$, (\ref{f1}) and (\ref{f2}), these restrictions become even milder, as the effective action at late times takes the form:
\bea\label{einstnd2}
S^{E}_{\mbox{\footnotesize late times}}&\sim&\int d^D x\sqrt{-g} ~\Bigg\{~ R  + e^{\frac{4\phi}{D-2}} \left[\frac{D-26}{6\ap} + f_0(T)\right] -\left[\frac{4(D-1)}{D-2} - 4f_2(T) \right]\partial \phi \cdot \partial \phi\nn
&&~~~~~~~~~~~~~~~~~ - f_3(T) \partial T \cdot \partial T  -f_4(T) \partial\phi\cdot \partial T ~\Bigg\} ~,
\eea
\nin We notice here that all the kinetic terms fall off as $t^{-2}$ at late times. Therefore, as long as $f_2$, $f_3$ and $f_4$ don't diverge faster than this, the kinetic terms for both fields effectively disappear from the action at late times, and one doesn't need to worry about ghost fields any more.

A final remark should be made about the tachyon potential, for which in our model we have considered the general form (in the Einstein frame):
\be
e^{\frac{4\phi}{D-2}} \left[f_1(T) \right]^{-\frac{D}{D-2}}\left[\frac{D-26}{6\ap}+f_0(T) \right]~.
\ee
For our interpolating solutions, as long as $f_0(T)$ does not diverge at late times, this falls asymptotically to zero (since $f_1\rightarrow \mbox{const}$ for late times, as can one see in figures \ref{fig1} and \ref{fig2}):
\be
e^{\frac{4\phi}{D-2}} \left[f_1(T) \right]^{-\frac{D}{D-2}}\left[\frac{D-26}{6\ap}+f_0(T) \right]\propto t^{-\frac{4}{D-2}} \left[\frac{D-26}{6\ap}+f_0(T) \right]~.
\ee
In the case of $f_1(T)=e^{-T}$, for our power-law expanding solutions, a similar relation is true as long as the tachyon amplitude is chosen to be greater than zero, $\tau_0>0$. Therefore, in all cases it seems that the presence of tachyons is not causing any instabilities to our model, within the constraints that we have placed before.

\section{Conclusions}
In this work we started from a configuration for a graviton, dilaton and tachyon background of a closed bosonic string, that is resummed in the Regge-slope parameter, $\ap$. After checking that this configuration is consistent with the conformal properties of the theory, to all orders in $\ap$, we studied the cosmological implications of this model. We found that in the sigma-model frame, this leads to a de Sitter universe, whereas in the Einstein frame there are two possible solutions: one is an inflationary universe, as long as a certain anti-alignment condition (\ref{antial}) is satisfied, and the other is a power-law expanding universe (that can also be chosen to be free of horizons), if the condition (\ref{antial}) is not satisfied. How the universe can exit the de Sitter era (which is characterized by cosmic horizons) and enter a power-law expanding era, by a way which disturbs the condition (\ref{antial}) is still unknown. However, we also found some interpolating solutions, i.e. solutions that contain an inflationary era, but interpolate between two flat universes (in the far past and in the far future), that are not related to the eternal inflation or power-law expanding solutions through a local field redefinition. These solutions are free of horizons, and since our approach is based on String scattering amplitudes, it is these solutions that are considered to be self-consistent configurations.
Throughout this work we considered our model to be in $D$ dimensions, and we showed that our configuration satisfies the conformal invariance conditions regardless of the value of $D$. Therefore, our work should be directly applied to $D=4$, without any need of compactification of extra dimensions.
As we mentioned in section \ref{exit}, particle production has been studied in universes which behave like (\ref{c1}) or (\ref{c1}) \cite{gubser}. It is reasonable to ask if string production is possible within our model, at the end of the inflationary era. The study of (p)reheating is, therefore, due to be done in our model. Finally, whether the model is phenomenologically realistic, with respect to its particle physics aspects, remains an open issue.

\section*{Acknowledgements}
I would like to thank J. Alexandre and N.E. Mavromatos for their contribution to this work. My work, as well as my participation to {\it Discrete '08} was supported by the European Union through the FP6 Marie Curie Research and Training Network UniverseNet (MRTN-CT-2006-035863).

\end{document}